\documentclass[12pt]{article}
\usepackage{epsfig}
\usepackage{rotating}
\newcommand{\gsim}{\lower.7ex\hbox{$\;\stackrel{\textstyle>}{\sim}\;$}}
\newcommand{\lsim}{\lower.7ex\hbox{$\;\stackrel{\textstyle<}{\sim}\;$}}
\newcommand{\bm}[1]{{\mbox{\boldmath $#1$}}}
\renewcommand{\AA}{{\bf A}\kern-1.5mm\raisebox{.7mm}{$\scriptstyle
\backslash$}}

\def\sidno{\ifcase\arabic{page}\or
  1\or 2\or 3\or 4\or \or 6\or 7\or 8\or 9\fi }

\begin{document}
\footnotesep=14pt
\begin{flushright}
\baselineskip=14pt
{\normalsize DAMTP-1999-181}\\
{\normalsize {astro-ph/0004098}}
\end{flushright}

\vspace*{.5cm}
\renewcommand{\thefootnote}{\fnsymbol{footnote}}
\setcounter{footnote}{0}
\begin{center}
{\Large\bf Cosmic Magnetic Fields From Particle 
Physics\footnote{Talk presented
at the 7th International Symposium on Particles, Strings and Cosmology
(PASCOS-99) at Granlibakken, Lake Tahoe, 10 - 16 Dec 1999,
to appear in the proceedings.}}
\end{center}
\setcounter{footnote}{2}
\begin{center}
\baselineskip=16pt
{\bf Ola T\"{o}rnkvist}\footnote{E-mail:
{\tt o.tornkvist@damtp.cam.ac.uk}}\\
\vspace{0.4cm}
{\em Department of Applied Mathematics and Theoretical Physics,}\\
{\em Centre for Mathematical Sciences, University of Cambridge,}\\
{\em Wilberforce Road, Cambridge CB3 0WA, United Kingdom}\\
\vspace*{0.25cm}{31 March 2000}
\end{center}
\baselineskip=20pt
\vspace*{.5cm}
\begin{quote}
\begin{center}
{\bf\large Abstract}
\end{center}
\vspace{0.2cm}
{\baselineskip=10pt%
I review a number of particle-physics models that lead to
the creation of magnetic fields in the early universe and address the
complex problem of evolving such primordial magnetic fields into
the fields observed today. Implications for future observations of the
Cosmic Microwave Background (CMB) are briefly discussed.}
\end{quote}
\renewcommand{\thefootnote}{\arabic{footnote}}
\setcounter{footnote}{0}
\newpage
\baselineskip=16pt
\section{Introduction to Cosmic Magnetic Fields}

Magnetic fields have been observed in a large
number of spiral galaxies, including the Milky Way \cite{Kronberg}.
In almost all cases,
the galactic field strengths are measured to be
a few times $10^{-6}$ G. This particular value
has also been found at a redshift of
$z=0.395$ \cite{redshift}
and between the galaxies in clusters.
In spiral galaxies, one discovers that the magnetic field is
aligned
with the spiral arms and density waves in the disk.
A plausible explanation is that
it
was
created by a mean-field dynamo mechanism \cite{dynamo}, in which a
much smaller seed
field was exponentially amplified by the turbulent motion of ionised gas
in conjunction with the differential rotation of the galaxy.

For the
dynamo to work, the initial seed
field must be correlated on a scale of $100$ pc, corresponding to the
largest turbulent eddy \cite{dynamo}.
The required strength of the seed field is subject to large uncertainties;
past authors have quoted
$10^{-21\pm 2}$ G as the lower bound
at the time of completed galaxy formation.
This would present a problem for most particle-physics and field-theory
inspired mechanisms of magnetic field generation. However, in recent
work with A.-C.\ Davis and M.J.\ Lilley \cite{bbounds}, I have shown that
the lower bound on the dynamo seed field can be significantly
relaxed if the universe is flat with a cosmological constant, as
is suggested by recent supernovae observations \cite{supernovae}.
In particular, for
the same dynamo parameters that give a lower bound of
$10^{-20}$ G for $\Omega_0=1$, $\Omega_\Lambda=0$,
we obtain
$10^{-30}$ G for $\Omega_0=0.2=1-\Omega_\Lambda$, implying that
particle-physics mechanisms could still be viable.
The observation at redshift $z=0.395$ \cite{redshift} can also be
accounted for with these parameters, but
requires a seed field of at least $10^{-23}$ G \cite{bbounds}.

The magnetic field is amplified by the 
dynamo until
its energy
reaches equipartition with the kinetic energy of
the
ionised gas, $\langle B^2/2\rangle=\langle\rho v^2/2\rangle$. 
Thus
a
final field of
$B_0\approx 10^{-6}$ G
results
for any seed field of sufficient strength.

It is difficult to explain the galactic field strength without
a dynamo mechanism. To this end, 
one would require a strong primordial field of
$10^{-3} (\Omega_0 h^2)^{1/3}$ G  at the epoch of radiation
decoupling $t_{\rm dec}$, corresponding to a
field
strength $10^{-9} (\Omega_0 h^2)^{1/3}$ G
on comoving scales of $1$ Mpc. Future precision measurements
of the CMB will put
severe constraints on such a primordial field \cite{CMB}.
Moreover,
magnetic fields on Mpc scales
have been probed by observations
of the Faraday rotation of polarised light from distant luminous
sources, which give an upper bound of about $10^{-9}$ G \cite{Kronberg}.
The observation of micro-Gauss fields between galaxies
in clusters presents an interesting dilemma. Because such regions are
considerably less dense than galaxies, it is doubtful whether a dynamo
could have been operative. Thus the intra-cluster magnetic fields, unless
somehow ejected from
galaxies, have formed directly from a primordial field stronger
than $10^{-3} (\Omega_0 h^2)^{1/3}$ G at $t_{\rm dec}$. Such a field would
certainly leave a signature in future CMB data \cite{CMB}.

Although particle-physics
inspired models typically produce weak
seed fields, they tend to
give precise predictions, unlike many
astrophysical mechanisms,
where the magnetic field strength is determined by complicated
nonlinear dynamics,
or solutions of general relativity with a magnetic field \cite{Thorne},
where the field strength must be fixed by observations.
With the possible exception of the last-mentioned model,
there is no compelling scenario that produces a
primordial
field strong enough to eliminate the need for
a dynamo.

Seed fields for the dynamo can be astrophysical or primordial.
In the former category there is
the important possibility that a seed field may arise spontaneously
due to non-parallel gradients
of pressure and charge density
during the collapse of a protogalaxy \cite{Kulsrud}. For the rest of
this talk, however, I shall assume that the seed field is primordial.

\section{Primordial Seed Fields}

It is
useful to
distinguish between primordial seed fields that are produced
with correlation
length smaller than
vs.\ larger than the horizon size.

{\sl Subhorizon-scale seed fields} typically arise
in first-order phase transitions and from causal processes involving
defects. For example, magnetic fields may be created on the surface
of bubble walls \cite{Hogan} due to local charge separation
induced by baryon-number gradients. The magnetic fields are then
amplified by plasma turbulence near the bubble wall. This possibility has been
explored for the QCD \cite{Cheng} as well as for the electroweak
\cite{baym-sigl} phase transition.

The production of magnetic fields in
collisions of expanding true-vacuum
bubbles will be discussed
in Sec.~\ref{bubs}. Fields can also be generated
in the wakes of, or due to the wiggles of, GUT-scale cosmic strings during
structure formation, resulting in a large correlation
length \cite{supercond}. Joyce and Shaposhnikov have shown that an
asymmetry of right-handed electrons, possibly generated at the GUT scale,
would become unstable to the generation of a hypercharge magnetic field
shortly before the electroweak phase transition \cite{Joyce}, leading to a
correlation length of order $10^{6}/T$.

{\sl Horizon-scale seed fields} emerge naturally in second-order phase
transitions of gauge theories
from the failure of
covariant derivatives of the Higgs field to correlate on superhorizon
scales \cite{Vacha}.

{\sl Superhorizon-scale seed fields} can arise as a solution of
the Einstein equations for axisymmetric universes \cite{Thorne} and
in inflationary or pre-Big Bang (superstring) scenarios. In the latter case,
vacuum fluctuations of the field tensor are amplified by the
dynamical
dilaton field \cite{Gasperini}. Inflationary models produce extremely
weak magnetic fields unless conformal invariance is explicitly
broken \cite{TurWid}, but even then great difficulties remain.
An exciting new possibility is that magnetic fields may be produced
via parametric resonance with an oscillating field \cite{Finelli}
e.g.\ during
preheating after inflation. 
Because the inflaton is initially coherent on
superhorizon scales, large correlations can arise without violating
causality.
A similar proposal involves 
charged scalar particles, minimally coupled to
gravity, that are created from the vacuum due to the changing
space-time geometry at the end of inflation. The particles give 
rise to fluctuating electric currents which are claimed to produce
superhorizon-scale (indeed, galactic-scale) fields of sufficient
strength to satisfy the galactic dynamo bound \cite{Calzetta}.
This mechanism deserves further investigation.

\section{The Evolution of Primordial Magnetic Fields}

Many
particle-physics and field-theory scenarios for producing primordial
magnetic fields result in too small a correlation length $\xi$,
which is a serious problem.
If the fields are produced
at the QCD phase transition or earlier with sub-horizon correlation
length, then the  expansion
of the universe cannot stretch
$\xi$ to more than 1 pc today (see Fig.~1).
This is far short of the
galactic dynamo lower bound of 5-10 kpc (comoving), corresponding
to 100 pc in a virialised galaxy \cite{bbounds}.

\begin{sidewaysfigure}
\epsfig{figure=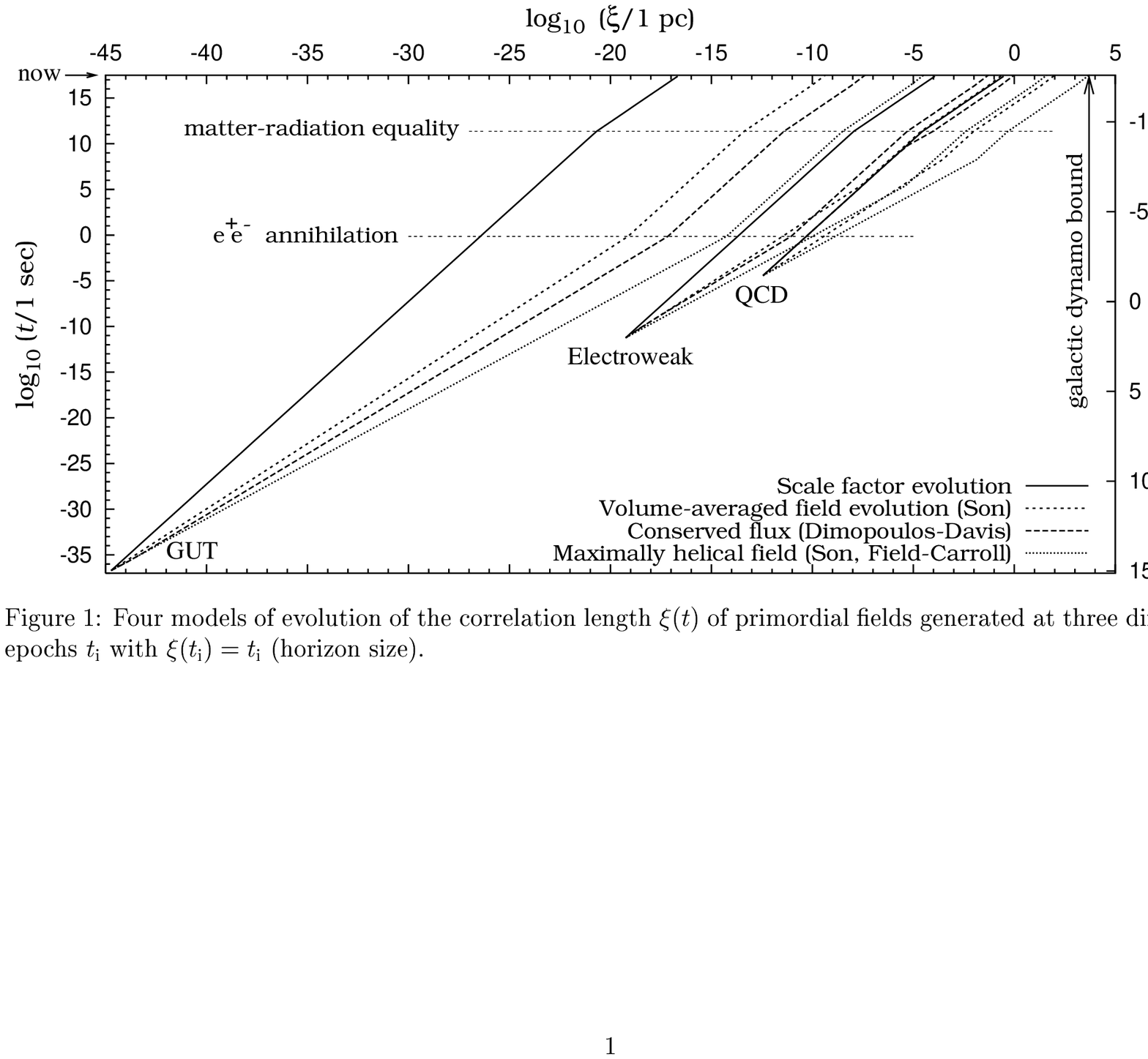,width=\textheight}
\end{sidewaysfigure}

Nevertheless, it has been argued by
many authors \cite{brandenburg,Son,carroll,kostas}
that the correlation length
will grow more rapidly due to magnetohydrodynamic (MHD) turbulence
and inverse cascade, which transfers power from
small-scale to large-scale Fourier modes. In Fig.~1, 
several non-relativistic
models for this
evolution are analysed. The most conservative estimate is
obtained by assuming that
the magnetic field strength on the scale of one correlation
length at any time
equals the volume average of fields that were produced on smaller
scales but have
since decayed \cite{Son}.
This leads to a growth $\xi\sim t^{7/10}$ (obtained from
the Minkowski-space growth $\xi\sim t^{2/5}$ via the substitution
$t\to \tau=t^{1/2}$ and multiplication by the scale factor). The most
optimistic estimate corresponds to the case
when the magnetic field has maximal
helicity in relation to the energy density \cite{Son,carroll}. As magnetic
helicity is approximately conserved
in the high-conductivity early-universe environment, one obtains
the growth law $\xi\sim t^{5/6}$.
Turbulence ends, freezing the growth (in comoving coordinates)
when the kinetic Reynolds number drops
below unity
at the $e^+e^-$ annihilation or later, depending on the model and
the parameters of the initial field.

An intermediate and rather plausible estimate has been given by
Dimopoulos and Davis \cite{kostas}, who use the fact that
the magnetic flux enclosed by a (sufficiently large) comoving closed
curve is conserved. The correlation length here increases at a rate
given by the Alfv\'{e}n velocity, so that $\xi\sim t^{3/4}$.

As Fig.~1 shows, only the most optimistic of these growth laws
leads to a correlation length today
that satisfies the galactic dynamo bound. This occurs
 for fields correlated over the horizon scale at the QCD phase transition.
Beware, however, that the growth laws were derived
using {\sl non-relativistic\/} MHD equations assuming
that the magnetic field energy density remains
in equipartition with the
kinetic energy density
$\rho\bar{v}^2/2$, where $\bar{v}$ is the presumed {\sl non-relativistic\/}
``bulk velocity'' of
the {\sl ultra-relativistic\/} plasma.
It seems plausible that a relativistic treatment could alter the predicted
evolution dramatically.
In this light, I
find it too early to reject the idea that also subhorizon fields
might evolve into fields sufficiently correlated to seed the galactic
dynamo.

At the same time, Fig.~1 demonstrates
the intrinsic advantage of
superhorizon field generation mechanisms.
For these,
the principal problem is not the correlation length,
 but to achieve sufficient strength of the magnetic field.

\section{Magnetic Fields From Bubble Collisions}
\label{bubs}

A simple solution for the magnetic field generated in a bubble
collision
was first obtained in a U(1) model
by Kibble and Vilenkin [KV] \cite{KibVil}.
We
have recently obtained an improved
analytical solution for the
field evolution in that model \cite{u1bub}
 using an analytical expression for the
bubble-wall profile.

Here I shall restrict attention
to {\em non-Abelian\/} bubble collisions, 
taking the minimal Standard Model as an example.
The initial Higgs field
in the two bubbles can be written
\begin{equation}
\Phi_1=\exp(-i\theta_0\bm{n}\cdot\bm{\sigma})
\left(\begin{array}{c}0\\*\rho_1(x)\end{array}\right)~,\quad\quad
\Phi_2=\exp(i\theta_0\bm{n}\cdot\bm{\sigma})
\left(\begin{array}{c}0\\*\rho_2(x)\end{array}\right)~\
\end{equation}
where $\bm{n}$, $\theta_0$ are constants.
As the bubbles collide, non-Abelian currents
$j_k^A=i[\Phi^\dagger T^A D_k\Phi - (D_k\Phi)^\dagger T^A \Phi]$
develop across the surface of their intersection, where
$T^A=(g'/2,g\sigma^a/2)$, $D_k=\partial_k-iW_k^A T^A$ and
$W_k^A=(Y_k,W_k^a)$. This, in turn, leads to
a ring-like flux of non-Abelian fields encircling the
bubble intersection. The recipe for projecting
out the electromagnetic field
amongst the non-Abelian fields in an
arbitrary gauge has been given elsewhere \cite{bdef}.

In the special cases $\bm{n}=(0,0,\pm 1)$ and $\bm{n}=(n_1,n_2,0)$
it is known \cite{SafCop} that the non-Abelian flux consists of
pure Z and pure W vector fields, respectively. 
When $n_3\neq 0,\pm1$, both $Z$ and $W$ fields are excited and
conspire to generate
 a non-zero electric current and associated magnetic field.
As is expected, the
resulting field strength is
of the order of $M_W^2/g$ with a correlation length $\xi\sim M_W^{-1}$.
However, when
plasma friction and conductivity are taken into account, the magnetic
field spreads over the interior of a bubble \cite{lilley}
leading to an appreciable increase in correlation length.

\subsection*{Acknowledgments}
The author is supported by the European Commission's TMR programme under
Contract No.~ERBFMBI-CT97-2697 and by the U.K.\ PPARC. He is grateful to
the Royal Society, {\em inter alii,} for travel support.

\end{document}